\documentclass[runningheads,a4paper]{llncs}

\usepackage{amssymb}
\usepackage{graphicx}
\usepackage{caption}

\usepackage{subcaption}
\usepackage{subfig}
\usepackage{algorithm}
\usepackage[noend]{algpseudocode}
\usepackage{algorithmicx}

\usepackage{url}
\urldef{\mailsa}\path|{pankaj.yadav, sriniwas.pandey, sraban}@iiitdmj.ac.in|    
\newcommand{\keywords}[1]{\par\addvspace\baselineskip
\noindent\keywordname\enspace\ignorespaces#1}

\begin{document}

\mainmatter  % start of an individual contribution

% first the title is needed
\title{Nearest Neighbor based Clustering Algorithm
	for Large Data Sets}

% a short form should be given in case it is too long for the running head
%\titlerunning{Lecture Notes in Computer Science: Authors' Instructions}

% the name(s) of the author(s) follow(s) next
%
% NB: Chinese authors should write their first names(s) in front of
% their surnames. This ensures that the names appear correctly in
% the running heads and the author index.
%
\author{Pankaj Kumar Yadav
\and Sriniwas Pandey \and Sraban Kumar Mohanty}
\authorrunning{Nearest Neighbor based Clustering Algorithms
	for Large Data Sets}
% (feature abused for this document to repeat the title also on left hand pages)

% the affiliations are given next; don't give your e-mail address
% unless you accept that it will be published
\institute{PDPM IIITDM JABALPUR, Madhya Pradesh, India 482005\\
\mailsa}

%
% NB: a more complex sample for affiliations and the mapping to the
% corresponding authors can be found in the file "llncs.dem"
% (search for the string "\mainmatter" where a contribution starts).
% "llncs.dem" accompanies the document class "llncs.cls".
%

%\toctitle{Lecture Notes in Computer Science}
%\tocauthor{Authors' Instructions}
\maketitle

\begin{abstract}

{ Clustering is an unsupervised learning technique in which data or objects are grouped into sets based on some similarity measure. Most of the clustering algorithms assume that the main memory is infinite and can accommodate the set of patterns. In reality many applications give rise to a large set of patterns which does not fit in the main memory. When the data set is too large, much of the data is stored in the secondary memory. Input/Outputs (I/O) from the disk are the major bottleneck in designing efficient clustering algorithms for large data sets. Different designing techniques have been used to design clustering algorithms for large data sets.
External memory algorithms are one class of algorithms which can be used for large data sets. These algorithms exploit the hierarchical memory structure of the computers by incorporating locality of reference directly in the algorithm. This paper makes some contribution towards designing clustering algorithms in the external memory model (Proposed by Aggarwal and Vitter 1988) to make the algorithms scalable.  
In this paper, it is shown that the {\it Shared near neighbors} algorithm is not very I/O efficient since the computational complexity is same as the I/O complexity. The algorithm is designed in the external memory model and I/O complexity is reduced. The computational complexity remains same.
We substantiate the theoretical analysis by showing the performance of the algorithms with their traditional counterpart by implementing in STXXL library.}
\end{abstract}

\section{Introduction}

Clustering is an unsupervised learning technique in which data or objects are grouped into sets based on some similarity measure. The data points in a group are similar and the points across the groups are dissimilar. There are few typical requirements for a good clustering technique in data mining \cite{zaiane2002,han2006}. Versatility, ability to discover clusters with different shapes, minimum number of input parameters, robustness with regard to noise, insensitive to the data input order, scalable to high dimensionality, scalable to large data sets are the important requirements. 

\subsection{Clustering of Large Data Sets}
The performance of the algorithm should not decrease with the increase of the data size. Most of the clustering algorithms are designed for small data sets and they fail to fulfill the last requirement i.e. scalable to large data sets. Many scientific, engineering and business applications frequently produces very large data sets~\cite{abello2002handbook}. The definition of ``Large'' varies with the changes in technology, mainly the memory and the computational speed of the computers. The data set which is large in today's computing environment may not remain as large after a few years. However the data size is increasing at much faster than the technology to handle it. Majority of the clustering algorithms are not designed to handle large data sets. There are a few approaches proposed in the literature to handle large data sets, e.g., Decomposition and Incremental approaches~\cite{Jain:1999,547613}. Parallel implementation is also used to handle large data sets.
\par Few algorithms are also devised in the literature which use preprocessing steps like summarization, incremental, approximation, distribution etc., to efficiently cluster large data sets. With the help of preprocessing steps, they actually store the summary of the data set and generate the clusters only considering the summary. Few examples include: Balanced Iterative Reducing and Clustering Using Hierarchies (BIRCH algorithm)~\cite{Zhang:1996}, CLARANS~\cite{1033770}, Clustering Using Representatives (CURE)~\cite{Guha:1998}, scalable K-mean++~\cite{Bahmani:2012}, etc. The BIRCH and CLARANS algorithms are suitable when the clusters are convex or spherical shape of uniform size. However, they compromise with the  quality when clusters have different sizes or non-spherical shapes~\cite{Zhang:1996}. Also random sampling and randomized search, are used by these algorithms which degrade the quality of the clustering because all the data points are not considered~\cite{zhang2010,0909-4412}.
\par In the traditional algorithm design, it is assumed that the main memory is infinite and it allows uniform and random access to all its locations. But in reality the present day computers have multiple levels of memory and accessing data from each level has its own cost and performance characteristics. If the data is too large to fit in the main memory then it has to be stored in the disk of the machine. Disk access time is millions times slower than the main memory access time. Most of the clustering algorithms assume that the main memory is large enough for the data set. However for large data sets, this is not a realistic assumption. So in case of large data, the usual computational cost may not be an appropriate performance metric but number of input/outputs (I/Os) can be more appropriate performance measure. Different designing techniques are used to design algorithms for large data sets. External memory algorithms  are one such class of algorithms which exploits the hierarchical memory structure of the computers by incorporating locality of reference directly in the algorithm~\cite{Aggarwal}. The external memory model was introduced by Aggarwal and Vitter in 1988. The Input/Output model (I/O-model) views the computer consisting of a processor, internal memory ($M$), and external memory (disk). The external memory is considered unlimited in size and is divided into blocks of $B$ consecutive data items. Transfer of a block of data between disk and RAM is called an I/O.

\subsection{Contribution of the paper}
Shared Near Neighbor  {\it SNN} ~\cite{1672233} is a technique, in which similarity of two points is defined based on the number of common neighbors, the two points share. The main advantage of the shared near neighbor based clustering algorithms is that the number of clusters is not specified, it is auto generated by the algorithm. Document clustering, temporal time series clustering are few examples where SNN clustering technique is used. In this paper, {\it SNN} based clustering algorithm is designed in external memory model to make it scalable. The computational as well as the I/O complexity of the SNN algorithm is $O(N^2k^2)$. This is the reason, why the SNN algorithm is not I/O efficient, hence unsuitable for large data sets. The traditional SNN algorithm is designed in the external Memory Model to make it I/O efficient. We show that the I/O complexity of the proposed algorithm is $O(N^2k^2/BM)$ which is a $BM$ factor improvement over the traditional SNN algorithm. The computational complexity remains same. Both traditional as well as proposed algorithms are implemented and the performance of the proposed algorithms is compared with its traditional counterpart. The proposed algorithm outperforms the in-core algorithm, as expected from the theoretical results.

\subsection{Organization of the paper}
This paper is organized as follows: In \textbf{Section 2}, the proposed scalable  shared near neighbors based clustering algorithm and its I/O analysis is described. \textbf{Section 3} contains the experimental results and observations.
The concluding remarks and future works are given in {\bf Section 4}.
\section{Proposed Scalable Clustering Algorithm based on SNN}

{\it Shared Near Neighbor (SNN)} is a technique in which similarity of two points is defined based on the number of neighbors, the two points share~\cite{1672233}. It can efficiently generate clusters of different sizes and shapes.
\par The inputs of the SNN algorithm are two parameters: $k$ (size of the nearest neighbors list) and $\theta$ (similarity threshold). The performance of the algorithm depends upon these parameters. In~\cite{moreira2013} an analytical process was proposed to find the most appropriate values of the input parameters.

\subsection{Traditional Shared Near Neighbors Algorithm}
The SNN algorithm has two steps. In the first step the k-nearest neighbor of all the points are calculated. Distance between two points can be calculated using any one of the distance measures. The k-nearest neighbors of a point are arranged in ascending order. As each point is its own zeroth neighbor so first point of each neighborhood row indicates the point number itself. In the second step the shared near neighbor of each data point is calculated. Assume that $i$ and $j$ with $i<j$ are any two points having at least $\theta$ (similarity threshold) matching neighbors and both points belong to each other's neighborhood list. Then the bigger index, e.g., $j$ is replaced by the smaller index, $i$. That means since $i$ and $j$ are similar so $j$ is labeled as $i$.
I/O complexity of this algorithm can be analyzed as follows: For the first step I/O complexity is $O(N^2D)$ and for second step number of I/Os is $O(N^2k^2)$. Hence overall complexity of the traditional algorithms is $O(N^2k^2)$ which is same as computational complexity.

\subsection{Proposed Scalable Shared Near Neighbors Algorithm}
The traditional algorithm is not very I/O efficient, hence it is not suitable for large data sets. In this section, we design the traditional algorithm in the external memory model to make it I/O efficient, hence scalable for large data sets. The computational steps of the proposed algorithm is same as the traditional algorithm but the data access pattern is modified to make the algorithm scalable. \\

\vspace{-.5cm}
\begin{algorithm}
	\caption{Proposed Algorithm For Generating the K-nearest neighbors matrix.}\label{abcde}
	%\scriptsize
	\begin{algorithmic}

		\State \textbf{Input:} Set of data points $ S \in {R}^{N\times D}$, i.e., $N$ points having dimension $D$ and $t$ is the block size.\\			
		\textbf{Output:} knn[N][k] // k-Nearest Neighbors Matrix.\\
		\hrule 
		\State {}
		{$ t={M}/{2(D+k)}$}
		\For {$i$=0 to $N/t-1$}
		\For{$j$=0 to $N/t-1$ }
		\State {\textbf{Read} $S_i$ and $S_j$ and also \textbf{Read} $knn_i$}
		\State {Do the following computations in main memory.}
		\For { $l=(i)t$ to $(i+1)t-1$}
		\For {$m=(j)t$ to $(i+1)t-1$}
		\State {sum = Distance$(l ,~ m)$}
		\If {dist[l][k] $>$ sum}
		{dist[l][k] = sum}
		\For {$d=k$ to 1 \& dist[l][d]$<$dist[l][d-1] }
		\State {Find the appropriate position in $ dist_{l,~d}$ 
			\State matrix block and also swap the index \State value of that point into the different \State matrix block $ knn_{l,~d}$}
		\EndFor
		\EndIf
		\EndFor
		
		\State {\textbf{Write} the matrix block $knn_{l,~d}$ into disk.}
		\EndFor
		\EndFor
		\EndFor
	\end{algorithmic}
\end{algorithm}\vspace{-1.0cm}
\subsubsection{Computation of k-Nearest Neighbor Matrix:}~
First step of the algorithm is I/O efficient generation of the k-nearest neighbor matrix. Assume that the $N\times D$ dataset is partitioned into $N/t$ blocks each of size $t\times D$. Here $t$ is a parameter to be fixed depending on the available main memory. Read any two blocks $S_i$ and $S_j$ into main memory and calculate the distance between each pair of points in the main memory. Store the distance in  a temporary  vector called ``dist'' of size $t\times k$ and corresponding points index in $knn$ matrix block of size $t\times k$. After computation of the k-nearest  neighbor of the block $S_i$, write the $knn$ matrix block into the external memory. Repeat the process for $N/t$ times. This process will generate $knn$ matrix and the procedure is described in Algorithm~\ref{abcde}. The main memory contains 2-blocks of size $t\times D$ and 2-blocks of size $t\times k$. Hence $M=2tD+2tk$.\vspace{-.5cm}
\subsubsection{The clustering step:~}
In the first step of the algorithm $knn$ matrix of size $N\times k$ is generated which is the input to the next phase of the algorithm. Assume that the matrix is divided into $N/t$ blocks of size $t\times k$ each.
Also assume that label table of size $N$ is divided into $N/t$ blocks of size $t$ each. Read any two blocks $knn_i$ and $knn_j$ where $i\le j$ and  also read two blocks of label table, $label_i$ and $label_j$ into the main memory. Then find all possible pair points satisfying the SNN similarity criteria  of $knn_i$ and  $knn_j$ blocks. In this way the label of all the points of the $knn_j$ block is calculated. Repeat the process for $N/t$ times. This process will generate cluster labels. The procedure is described in Algorithm~\ref{abed}.

\begin{figure}[!ht]
	\centering
	\includegraphics[width=\textwidth,height=.6\textwidth]{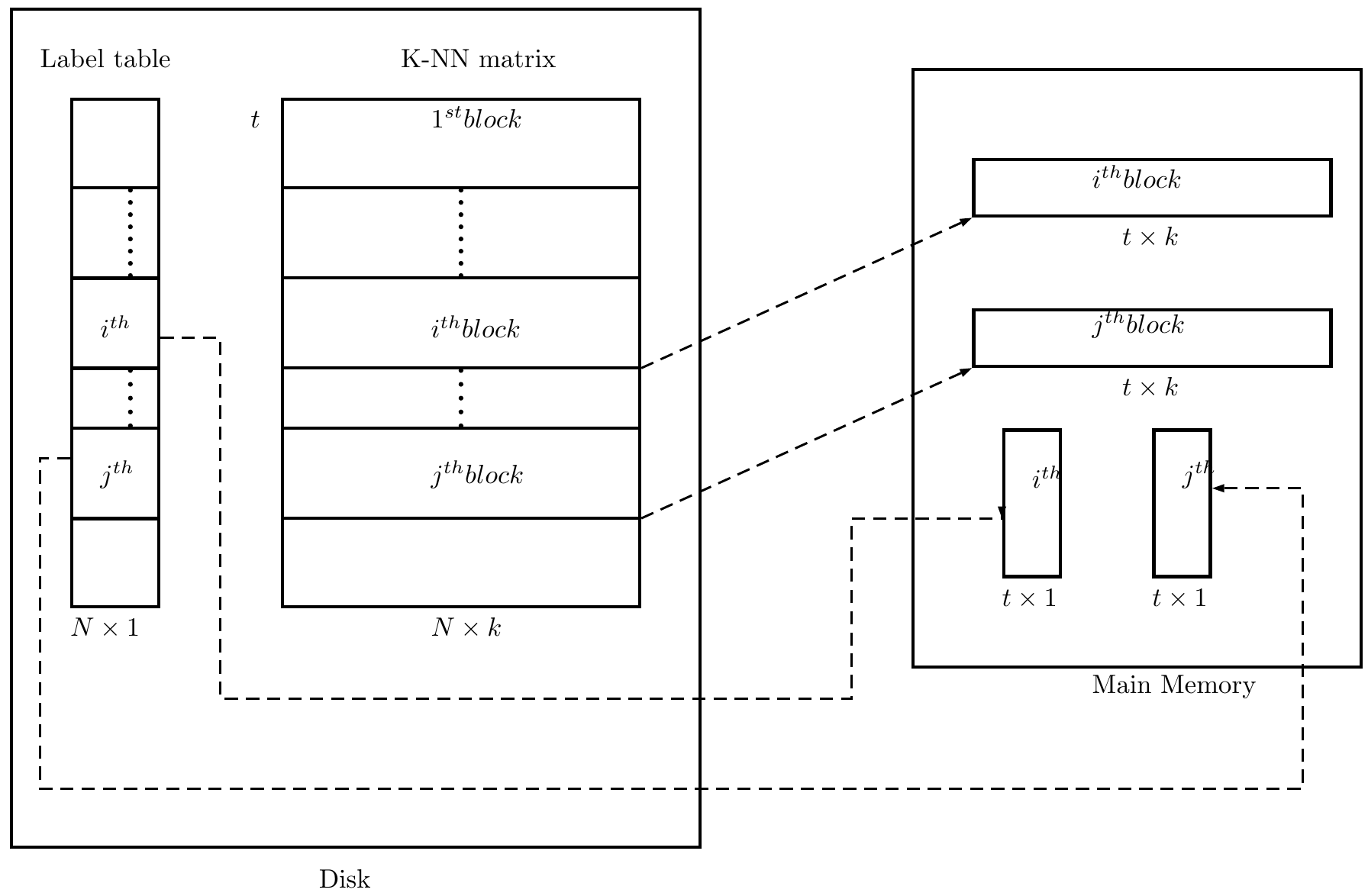}
	\caption {Snapshot of the transfer of blocks between disk and main memory}
	\label{algo}
\end{figure}

Here the main memory contains 2-blocks of size $t\times k$ and 2-blocks of size $t$. Hence $M=2tk+2t$.  The transfer of blocks between main memory and disk is shown Figure \ref{algo}.

%\newpage

\begin{algorithm}[!ht]
	\caption{ Proposed Shared Near Neighbors Algorithm}\label{abed}
	%\scriptsize
	\begin{algorithmic}
		
		\State\textbf {Input:} $~~~\theta$ $~~~~~~~~~~~~~~~$//Similarity threshold.
		\State \hspace*{10mm}{$~~~~knn[N][k]$} $~~~$//k-nearest neighbor matrix
		\State \hspace*{10mm}{$~~~~label[N]$} $~~~~~~~$//cluster label of each point which is initialized as $label[i]=i$\\
		
		\textbf {Output:} label[N] $~~~~~$//Cluster labels.\\
		\hrule
		\State {$t={M}/{k}$}
		\For {$i=0$ to $N/t-1 $}
		\For {$j=i$ to $N/t-1$}
		\State {\textbf{Read} matrix block $knn_i$ and $knn_j$}
		\State {Do the following computations in main memory}
		\For {$r=(i)t$ to $(i+1)t-1$}
		\For {$l=(i)t$ to $(i+1)t-1$}
		\For {$m=0$ to $k$}
		\If {$knn[r][0]$==$knn[l][m]$} {count=1}
		\EndIf
		\EndFor
		\For {$m=0$ to $k$}
		\If {$knn[l][0]$==$knn[r][m]$} {count++}
		\EndIf
		\If {$count$==2}
		\For {$p=1$ to $k$}
		\For {$q=1$ to $k$}
		\If {$knn[r][p]==knn[l][q]$} {n++}
		\EndIf
		\EndFor
		\EndFor
		\EndIf
		\If {$n>\theta$}
		\If {$i==j$}
		\State {\textbf{Read} the block $label_i$ of size $1 \times t$ into main memory}
		\If {$label[r]>label[l]$}
		\State {$label[r]=label[l]$}
		\Else 
		\State {$label[l]=label[r]$}
		\EndIf
		\Else
		\State {\textbf{Read} the block $label_i$ of size $1 \times t$ into main memory}
		\If {$label[r]>label[l]$}
		\State {$label[r]=label[l]$}
		\Else 
		\State {$label[l]=label[r]$}
		\EndIf
		\EndIf
		\EndIf
		\EndFor
		\EndFor
		\EndFor
		\State {$count=0,n=0$}
		\EndFor
		
		\EndFor
		
	\end{algorithmic}
\end{algorithm}

\subsection{I/O Analysis}
Traditional algorithm takes $O(N^2k^2)$ number of I/Os. I/O complexity of the proposed algorithm is described here. 

\subsubsection{ I/O complexity of Algorithm~\ref{abcde}} %\vspace*{-.3cm}
The main memory contains 2-blocks of size $t\times D$ and 2-blocks of size $t\times k$. Hence $M=2tD+2tk$, i.e., $t=\Theta(M/(D+k))$. Total number of I/Os required to generate the $knn$ matrix is = $O((ND/B)N/t)=O((N^2D/B)((D+k)/M))=O(({N^2Dk}+{N^2D^2})/{BM})$ % \vspace{-0.35cm}.

\subsubsection{ I/O complexity of Algorithm \ref{abed}} %\vspace*{-.3cm}
Here $M=2tk+2t$, i.e., $t=\Theta(M/k)$.
Total number of I/Os required by the algorithm is= $O(((Nk+N)/B)N/t)=O((N^2k+N^2)/tB)=O({N^2k^2}/{BM})$ \vspace{0.25cm}.

So the total number of I/Os incurs by two phases of the algorithm is $O(({N^2Dk}+{N^2D^2}+N^2k^2)/{BM})$. The dimension $D$ is a constant, so ignoring the constant term , the I/O complexity of the algorithm is $O(N^2k^2)/{BM})$ which is a $BM$ factor improvement over the traditional algorithm.
%	\clearpage
\section{Experimental Results} %\vspace*{-0.0cm}
\subsection{Performance of the Proposed Algorithm}
Many external memory software libraries are being designed. Few of them to mention are STXXL~\cite{2008}, LEDA-SM~\cite{crauser1999}, TPIE~\cite{arge2002}. STXXL is used in our implementation. STXXL is the implementation or adaptation of C++ STL (standard template library) for external memory computations~\cite{cc2009}. Both the traditional and the proposed algorithms are implemented in STXXL~\cite{2008}. 
\par Since both the algorithms follow exactly same computational steps, computational complexity remains same and so both of them generate same set of clusters. Hence the quality analysis is omitted. Our main focus is on analyzing  and reducing the I/O complexity of algorithm. 
\par The data sets are generated randomly. The dimension of the data is $64$ and the size of the data set varies from $10000~ to~ 200000$. The main memory size is restricted to 1 MB only and the Hard disk size is $150$ GB. The algorithm is implemented on  Ubuntu $12.04$ system with a $2.0$ GHz CPU(Intel Core 2 Duo) and $2$ GB main memory. 

\begin{figure}[!ht]
	
	\begin{minipage}{.48\linewidth}
		\includegraphics[height=\textwidth,width=1.2\textwidth]{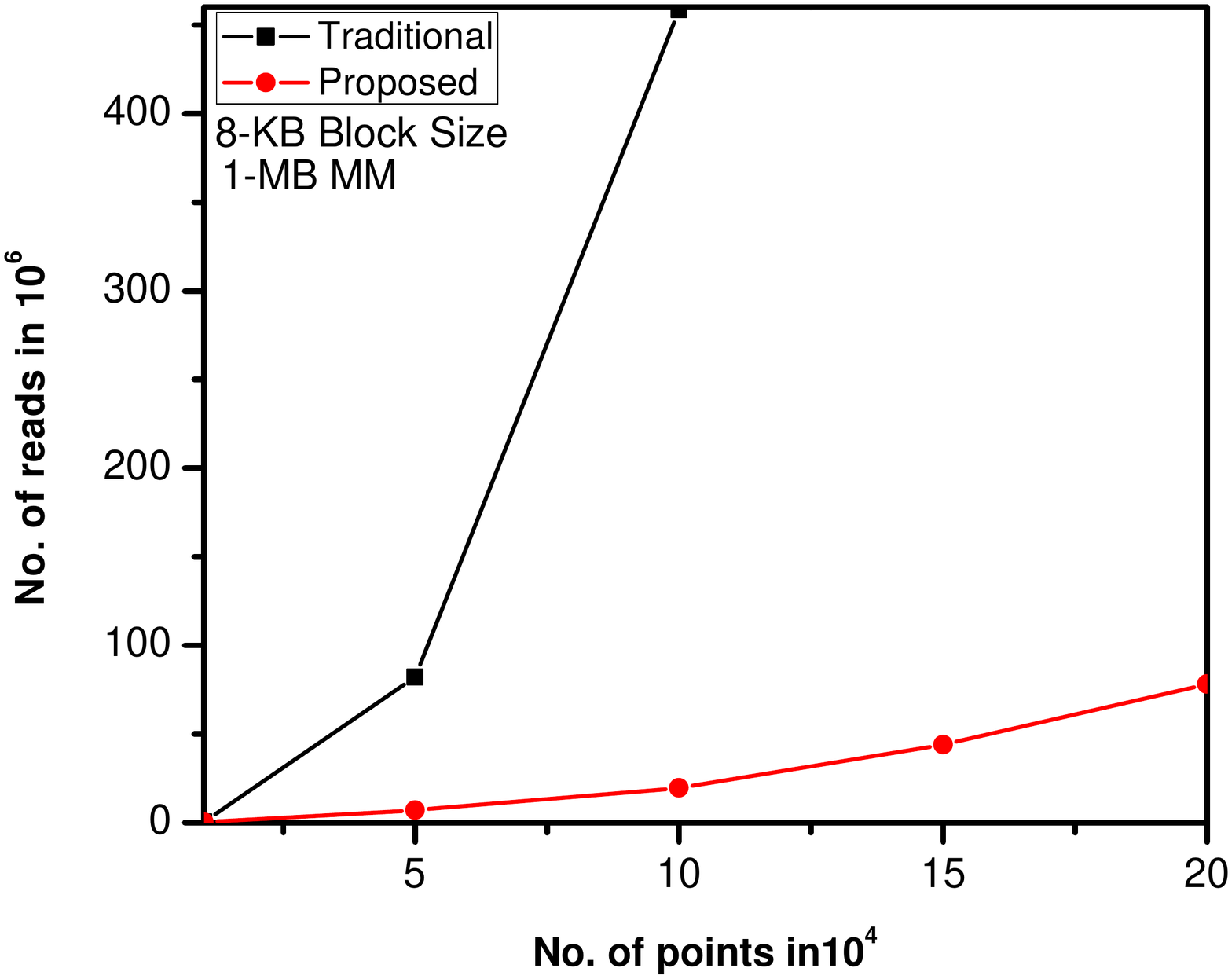}
		\subcaption{Number of reads}
		\label{zab}
	\end{minipage}
	\begin{minipage}{.48\linewidth}
		\includegraphics[height=\textwidth,width=1.2\textwidth]{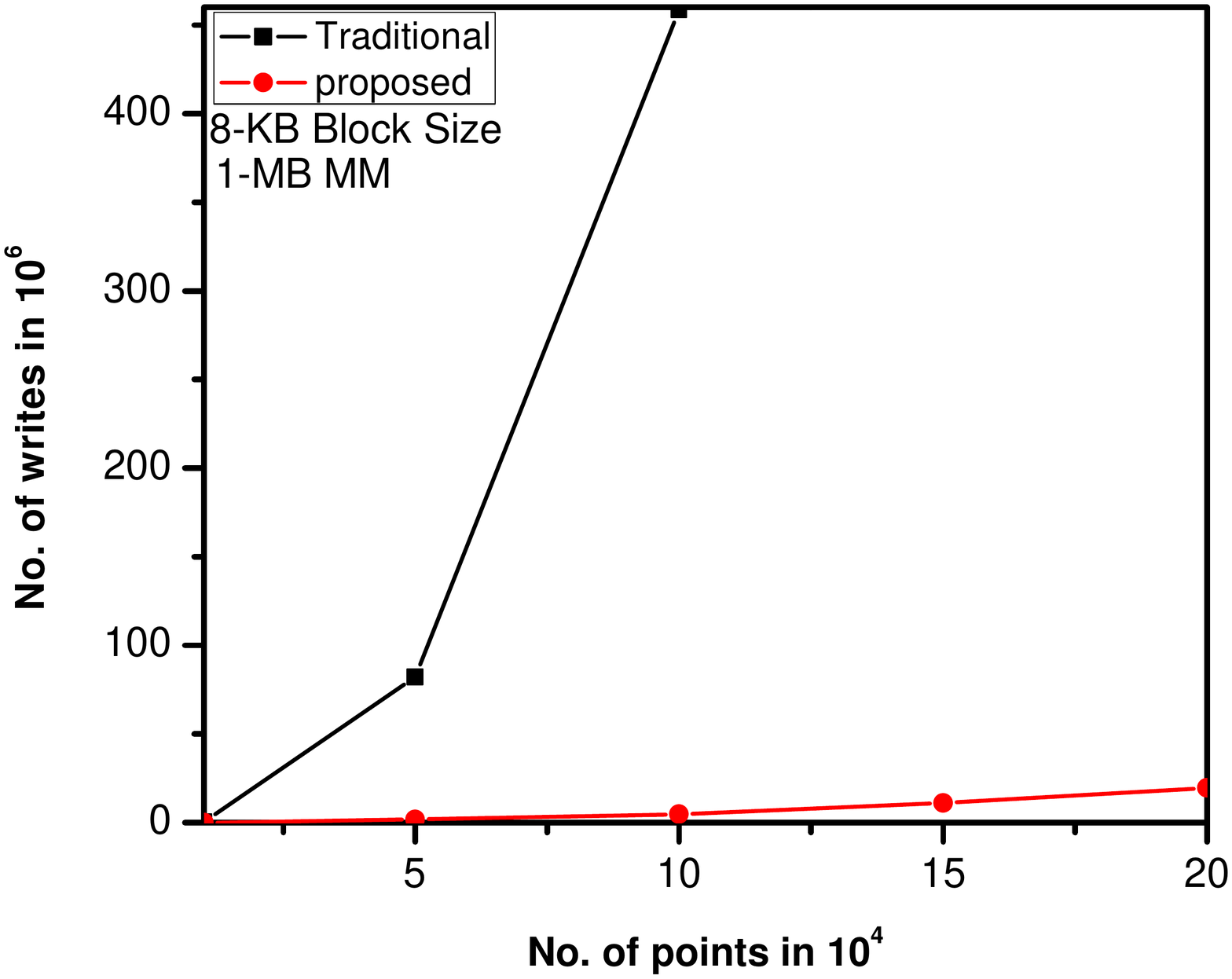}
		\subcaption{Number of writes}
		\label{zbb}
	\end{minipage}

	\begin{minipage}{.48\linewidth}
		\includegraphics[height=\textwidth,width=1.2\textwidth]{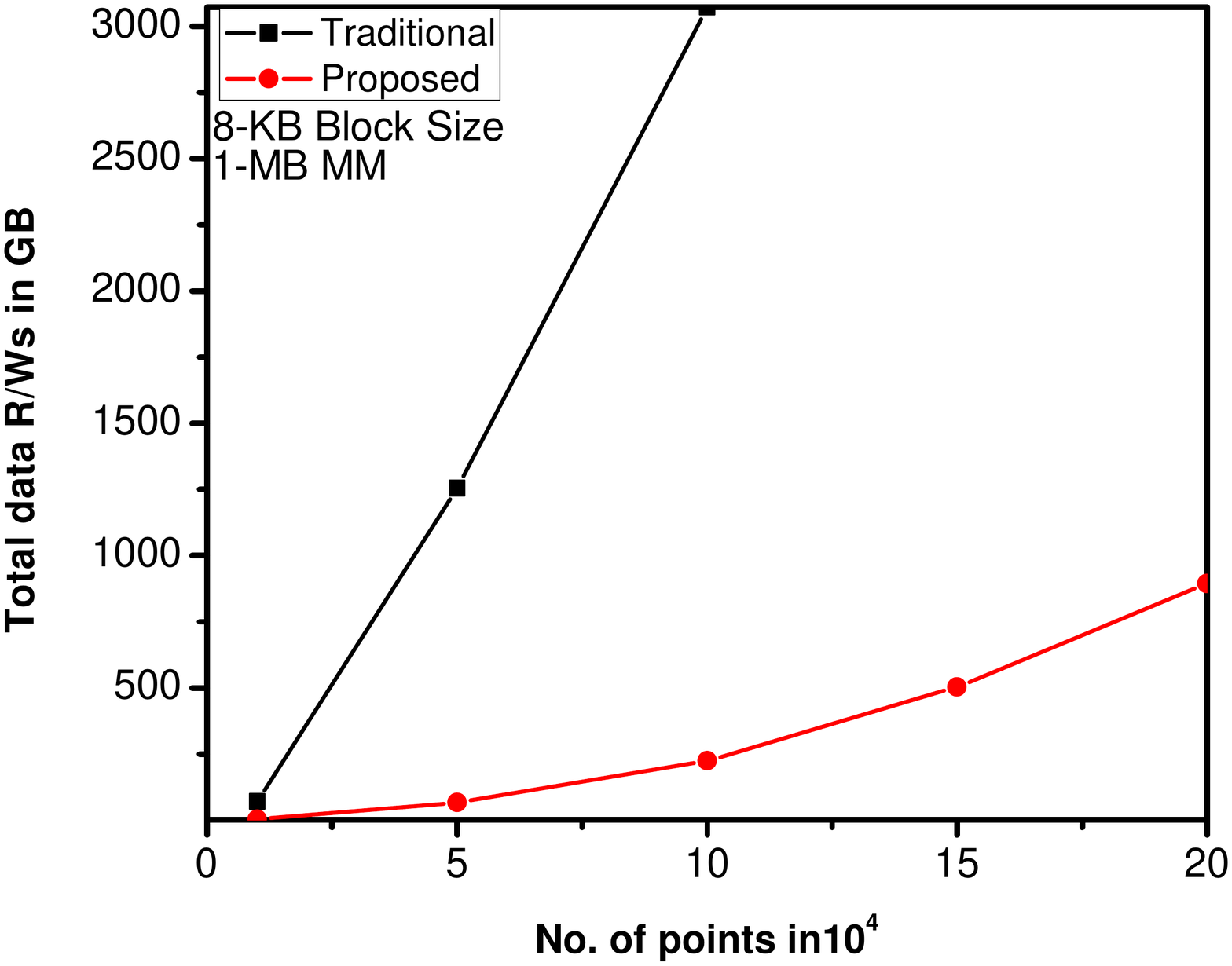}
		\subcaption{Total Data read/write in GBs}
		\label{zcb}
	\end{minipage}
	\begin{minipage}{.48\linewidth}
		\includegraphics[height=\textwidth,width=1.2\textwidth]{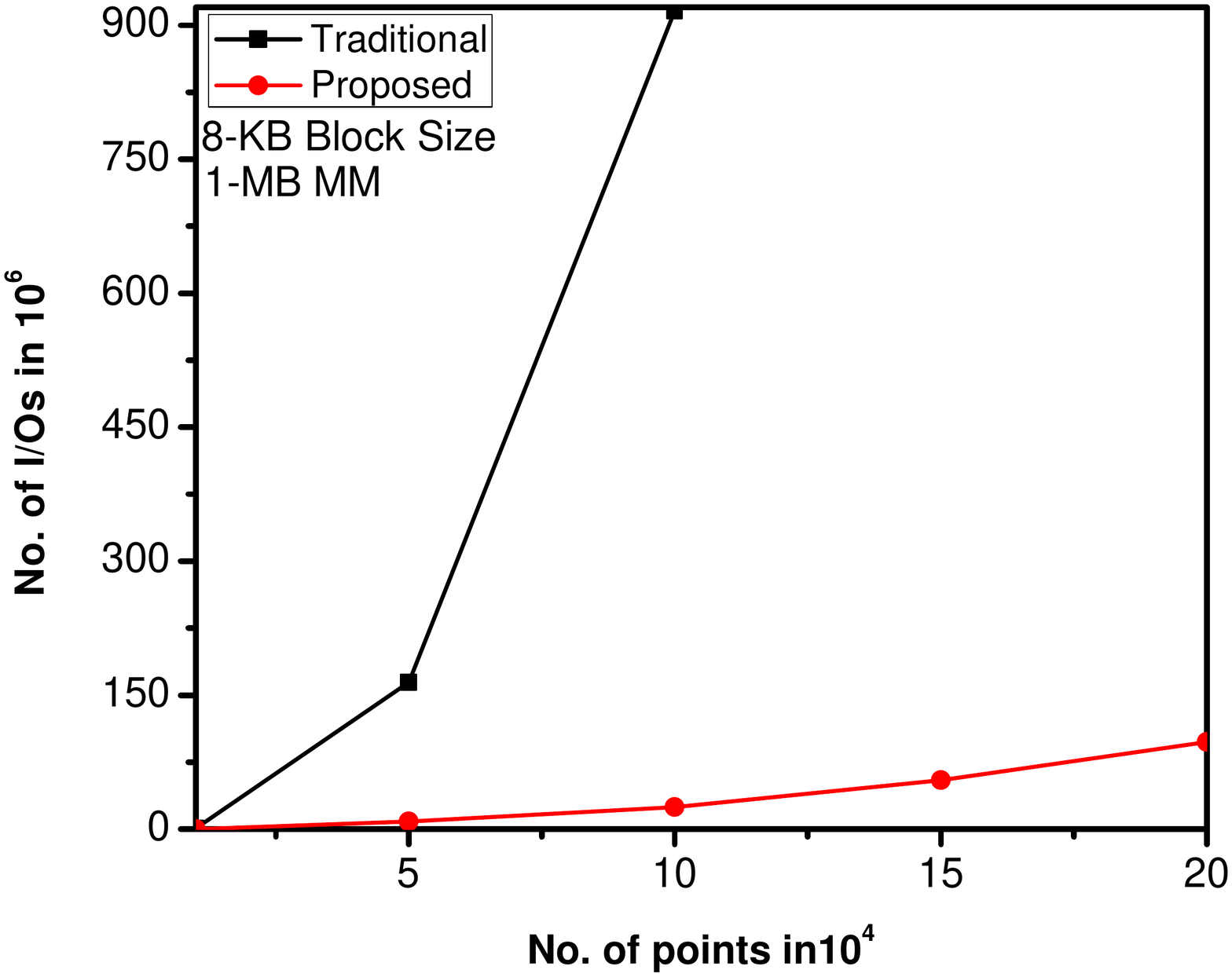}
		\subcaption{Total number of I/Os}
		\label{zdb}
	\end{minipage}
	%	\centering
	\caption{Performance of Proposed algorithm}
\end{figure}

For ease of implementation the algorithmic block size ($t$) is set same as the disk block size ($B$). When the block size is set to 8 KB and available main memory is restricted to 1 MB total number of reads or writes goes beyond 500$\times 10^6$ for 10$\times10^4$ data points in case of traditional algorithm. While in proposed algorithm number of read or writes is less than $100\times 10^6$ even for $20\times 10^4$ data points. The similar results were obtained for total number of I/Os and total data read and written. Total number of I/Os for traditional algorithm exceeds $900\times10^6$ for $10\times10^4$ data points while it is less than $150 \times 10^6$ for $20\times 10^4$ data points in case of proposed algorithm. In-core algorithm fails to give result after $5$ days for $15\times 10^4$ points. Figure \ref{zab}, \ref{zbb}, \ref{zcb} and \ref{zdb} illustrate number of reads, writes, data R/Ws and I/Os respectively. \\

%\begin{figure}[H]
%	\begin{minipage}{.20\linewidth}
%	\centering
%	
%		\includegraphics[height=\textwidth]{./image/chapt-3/snn/Graph1}
%	\subcaption{Number of reads}
%	\label{mna}
%\end{minipage}
%	\begin{minipage}{.20\linewidth}
%	\includegraphics[height=\textwidth]{./image/chapt-3/snn/Graph2}
%	\subcaption{Number of writes}
%	\label{mnb}
%
%\end{minipage}
%
%\begin{minipage}{.20\linewidth}
%	
%	\includegraphics[height=\textwidth]{./image/chapt-3/snn/Graph3}
%	\subcaption{Total data R/Ws in GB}
%	\label{mnc}
%\end{minipage}
%
%%\begin{minipage}{.20\linewidth}
%%
%%
%%	\includegraphics[height=\textwidth]{./image/chapt-3/snn/Graph4}
%%	
%%	\subcaption{Total number of I/Os}
%%	\label{mnd}%\label{dd}
%%\end{minipage}
%\end{figure}
\subsection{Effect of Main Memory Size on the Performance of the Proposed Algorithm}
The proposed algorithm is run on different sizes of main memory to study the effect of main memory on the performance of the algorithm. The main memory is restricted to 1MB, 4MB, 16MB and 128MB.
\par
\begin{figure}[!ht]

	\begin{minipage}{.32\linewidth}
		\includegraphics[height=\textwidth,width=1.2\textwidth]{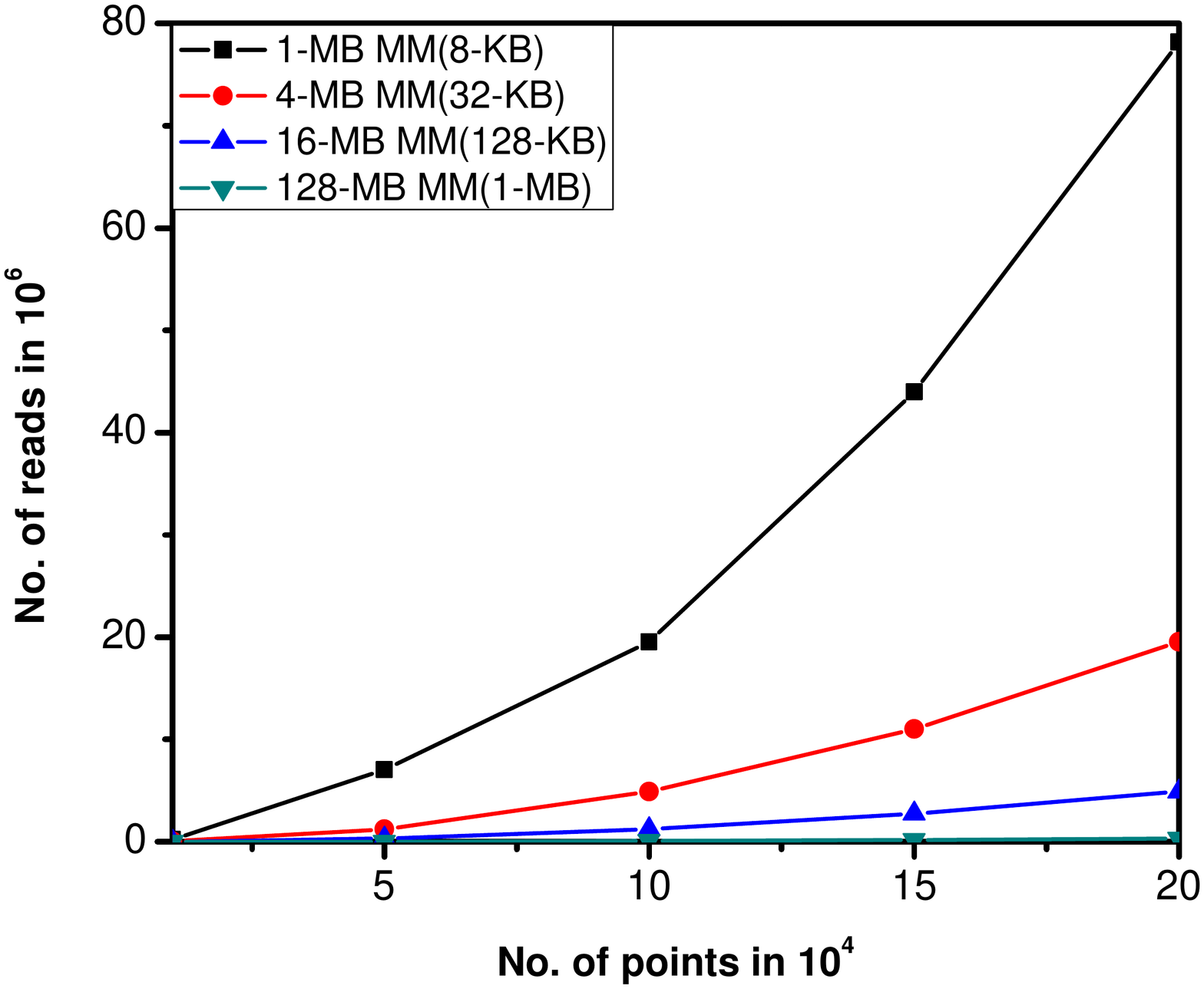}
		\subcaption{Number of reads}
		\label{kpa}
	\end{minipage}
	\begin{minipage}{.32\linewidth}
		\includegraphics[height=\textwidth,width=1.2\textwidth]{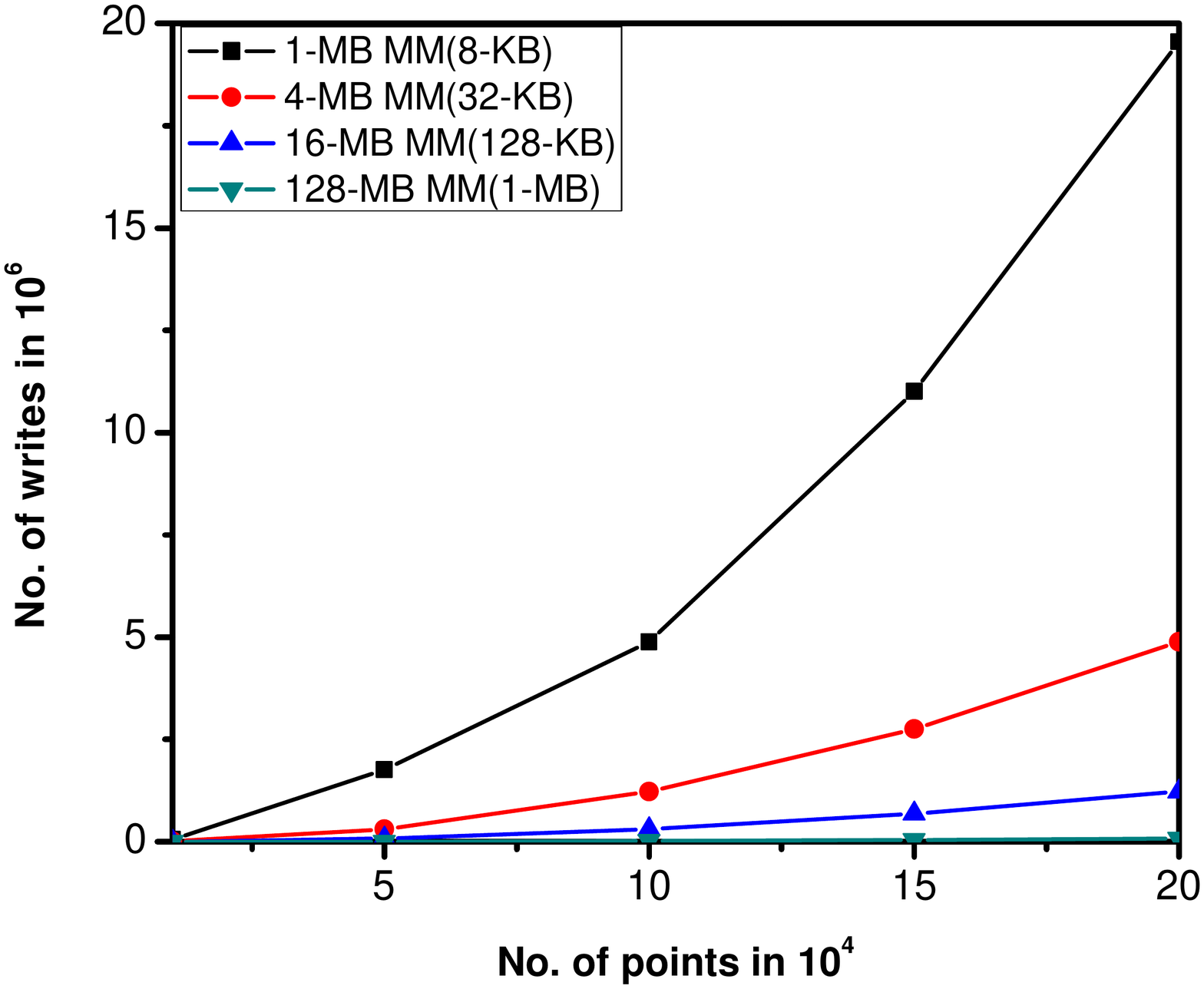}
		\subcaption{Number of writes}
		\label{kpb}
	\end{minipage}
	\begin{minipage}{.32\linewidth}
		\includegraphics[height=\textwidth,width=1.2\textwidth]{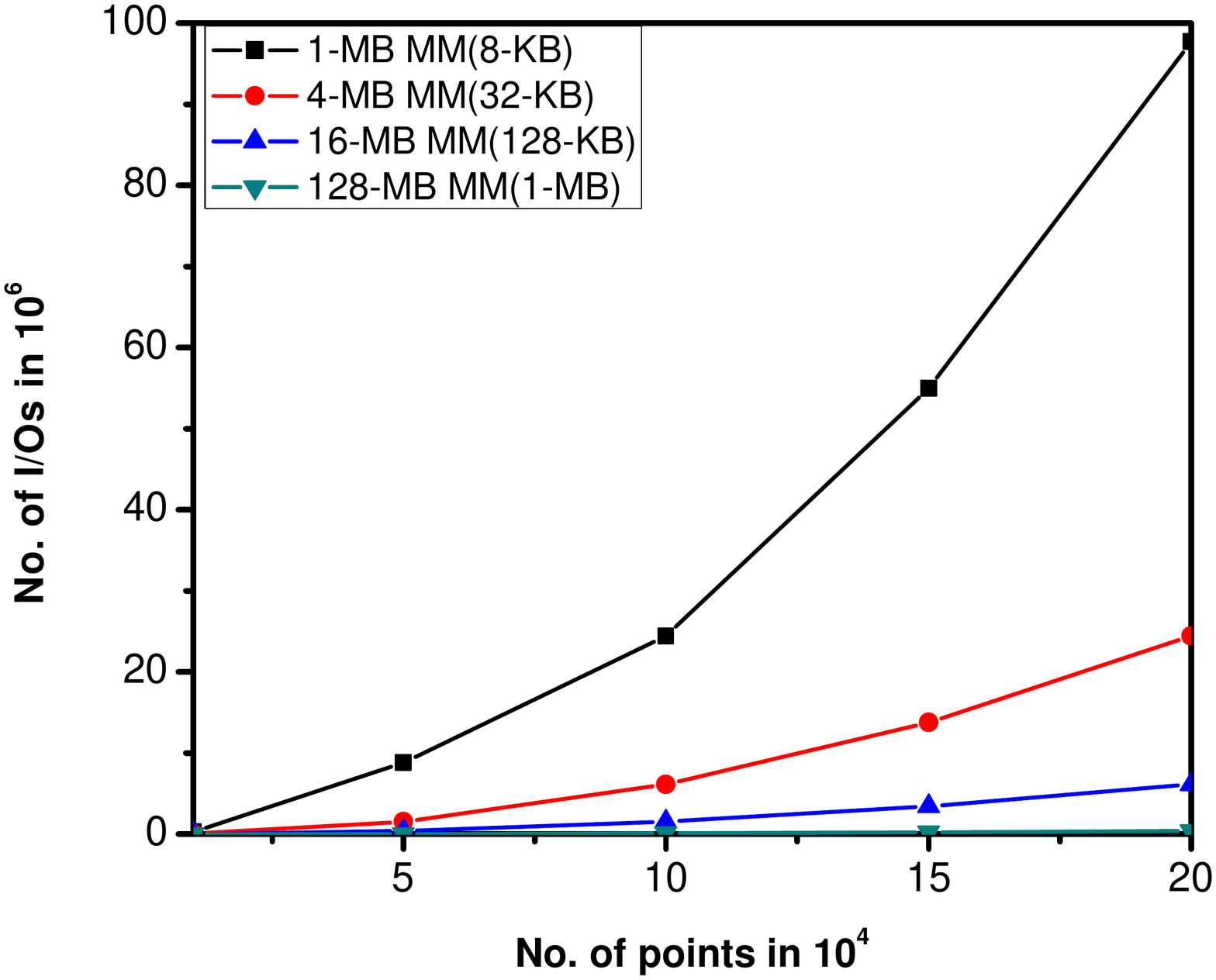}
		\subcaption{Total number of I/Os}
		\label{kpc}
	\end{minipage}
		\caption{Effect of main Memory Size}	
\end{figure}

 It is clear from the graph that the I/O reduces as the main memory increases. If we closely observe a graph we can see that when the number of data points is $20\times 10^4(\approx 100MB)$ and main memory size is 128 MB, the line denoting the total number of I/O is very close to x-axis. Similar effect of main memory size can be seen in other graphs as well.
 That substantiated the theoretical I/O analysis as the I/O is dependent on available main memory size.  Figure  \ref{kpa}, \ref{kpb} and \ref{kpc} illustrate number of reads, writes and I/Os respectively.

%\begin{figure}[h]
%	\centering
%	\includegraphics[width=8cm,height=6cm]{./image/chapt-2/Graph1.pdf}
%	\caption{Number of reads}
%	\label{kpa}
%\end{figure}
%
%\begin{figure}[H]
%	\includegraphics[width=8cm,height=6cm]{./image/chapt-2/Graph2.pdf}
%	\caption{Number of writes}
%	\label{kpb}
%\end{figure}
%
%\begin{figure}[H]
%	\centering
%	
%	\includegraphics[width=8cm,height=6cm]{./image/chapt-2/Graph3.pdf}
%	\caption{Total number of I/Os}
%	\label{kpc}
%\end{figure}

\section{Conclusion}
\thispagestyle{empty}
This paper makes some contribution in the field of big data clustering by redesigning the existing algorithm in external memory model. The shared near neighbors (SNN) algorithm has been designed on external memory model. It is shown that the I/O complexity of
the proposed algorithm is $O(N^2k^2/BM)$ which is a $BM$ factor improvement over the traditional SNN algorithm. Both algorithms produce the same set of clusters. Both algorithms are implemented in STXXL to compare their performance with the in-core algorithms.
The design technique can be used to adapt various existing algorithms for large data.
Without theoretical analysis it is often difficult to say which clustering algorithm will perform better for different sized data sets. So one of our future work is to analyze the I/O complexity of the best known clustering algorithms of the literature and design them on the external memory model to make them suitable for massive data sets.
\bibliographystyle{splncs03} 
\bibliography{reference}

\begin{thebibliography}{10}
\providecommand{\url}[1]{\texttt{#1}}
\providecommand{\urlprefix}{URL }

\bibitem{abello2002handbook}
Abello, J., Pardalos, P.M., Resende, M.G.: Handbook of massive data sets.
  Springer (2002)

\bibitem{Aggarwal}
Aggarwal, A., Vitter, J.: The input/output complexity of sorting and related
  problems. Communications of the ACM  31(9),  1116--1127 (1988)

\bibitem{arge2002}
Arge, L., Procopiuc, O., Vitter, J.: Implementing {I/O}-efficient data
  structures using {TPIE}. In: Algorithms — ESA 2002, pp. 88--100. Springer
  (2002)

\bibitem{Bahmani:2012}
Bahmani, B., Moseley, B., Vattani, A., Kumar, R., Vassilvitskii, S.: Scalable
  k-means++. Proceedings of the VLDB Endowment  5(7),  622--633 (2012)

\bibitem{crauser1999}
Crauser, A., Mehlhorn, K.: {LEDA-SM: Extending LEDA} to secondary memory. In:
  Algorithm Engineering, pp. 228--242. Springer (1999)

\bibitem{2008}
Dementiev, R., Kettner, L., Sanders, P.: {STXXL: standard template library for
  XXL }data sets. Software: Practice and Experience  38(6),  589--637 (2008)

\bibitem{0909-4412}
El-Sharkawi, M.E., El-Zawawy, M.A.: Algorithm for spatial clustering with
  obstacles. In: International Conference on Intelligent Computing and
  Information Systems (ICICIS’02) (2002)

\bibitem{Guha:1998}
Guha, S., Rastogi, R., Shim, K.: {CURE: }an efficient clustering algorithm for
  large databases. In: ACM SIGMOD Record. vol.~27, pp. 73--84. ACM (1998)

\bibitem{han2006}
Han, J., Kamber, M.: Data Mining, Southeast Asia Edition: Concepts and
  Techniques. Morgan kaufmann (2006)

\bibitem{Jain:1999}
Jain, A.K., Murty, M.N., Flynn, P.J.: Data clustering: A review. ACM computing
  surveys (CSUR)  31(3),  264--323 (1999)

\bibitem{1672233}
Jarvis, R.A., Patrick, E.A.: Clustering using a similarity measure based on
  shared near neighbors. IEEE Transactions on Computers  100(11),  1025--1034
  (1973)

\bibitem{547613}
Judd, D., McKinley, P.K., Jain, A.K.: Large-scale parallel data clustering. In:
  Proceedings of the 13th International Conference on Pattern Recognition.
  vol.~4, pp. 488--493. IEEE (1996)

\bibitem{moreira2013}
Moreira, G., Santos, M.Y., Moura-Pires, J.: {SNN Input Parameters}: how are
  they related{?} In: International Conference on Parallel and Distributed
  Systems (ICPADS). pp. 492--497. IEEE (2013)

\bibitem{cc2009}
Musser, D.R., Derge, G.J., Saini, A.: STL tutorial and reference guide: C++
  programming with the standard template library. Addison-Wesley Professional
  (2009)

\bibitem{1033770}
Ng, R.T., Jiawei, H.: {CLARANS:} a method for clustering objects for spatial
  data mining. IEEE Transactions on Knowledge and Data Engineering  14(5),
  1003--1016 (2002)

\bibitem{zaiane2002}
Za{\"\i}ane, O.R., Foss, A., Lee, C.H., Wang, W.: On data clustering analysis:
  Scalability, constraints, and validation. In: Advances in Knowledge Discovery
  and Data Mining, pp. 28--39. Springer (2002)

\bibitem{Zhang:1996}
Zhang, T., Ramakrishnan, R., Livny, M.: {BIRCH:} an efficient data clustering
  method for very large databases. In: ACM SIGMOD Record. vol.~25, pp.
  103--114. ACM (1996)

\bibitem{zhang2010}
Zhang, X.: Contributions to Large Scale Data Clustering and Streaming with
  Affinity Propagation. Application to Autonomic Grids. Ph.D. thesis,
  Universit{\'e} Paris-Sud (2010)

\end{thebibliography}

\end{document}